\documentclass[num-refs]{wiley-article}

\usepackage{siunitx}
\usepackage{lineno,hyperref}
\usepackage[utf8]{inputenc}
\usepackage[english]{babel}
\usepackage[T1]{fontenc}
\usepackage{amsmath}
\usepackage{tikz}
\usepackage{lipsum}
\usepackage{float}

\papertype{Original Article}
\paperfield{Journal Section}

\title{Nonlinear LC Circuit with Josephson Junction}

\author[1]{Alberto Delgado PhD}

\affil[1]{Electrical and Electronics Engineering Department, National University of Colombia, Bogota, 11001, Colombia}

\corraddress{Alberto Delgado PhD, Electrical and Electronics Engineering Department, National University of Colombia, Bogota, 11001, Colombia}
\corremail{adelgado@ieee.org}

\fundinginfo{National University of Colombia}

\runningauthor{Alberto Delgado}

\begin{document}

\maketitle

\begin{abstract}
This paper illustrates a unified approach, classical circuit and control theories, to study a nonlinear LC circuit with a current dependent inductance as model of the Josephson junction, the mathematical analysis is complemented with simulations to visualize different dynamics under changes in parameters and inputs. Currently, in quantum computing, superconducting circuits with Josephson junctions are used as the building blocks of quantum bits or qubits.

\keywords{feedback linearization, \emph{Josephson junction}, linear state feedback,  machine learning, \emph{nonlinear circuits}, Taylor linearization}
\end{abstract}

\section{Introduction}
Recently there has been a growing interest in quantum computing hardware using superconducting circuits as quantum bits or qubits, the main component in these circuits is the Josephson junction \cite{devoret, wendin} used to improve the energy resolution between ground and excited states. On the other hand, in classic circuit theory, a Josephson junction can be modelled as a nonlinear inductance with possible applications in tunable filters \cite{zhou}. Also within a unified classical approach, circuit theory and control theory, the dynamics of a Josephson junction has been studied in magnetically coupled circuits \cite{delgado} using the concepts of state variables and the phase plane as part of the analysis, simulations are included to visualize the mathematical results. The main goals of this paper are: (i) to introduce the Josephson junction as another circuit element; (ii) study in detail a nonlinear LC oscillator, with Josephson junction, using a unified approach of circuit and control theories including an application of machine learning with a neural network.

The paper is organized as follows. Section two introduces the LC circuit state equations with a nonlinear inductance that models the Josephson junction. Section three is the corresponding Taylor linearization of the nonlinear circuit around an operating point ${(\bar{x}_1, \bar{x}_2, \bar{u})}$. Section four presents the second order transfer function, from the Taylor linearization, with a resonance frequency that depends on the operating point. Section five is an alternative to Taylor linearization known as input - output linearization where the nonlinear terms are cancelled exactly to obtain a linear dynamics between a new circuit input and the given output. Section six is a practical implementation of nonlinear feedback with a neural network to overcome, for this circuit with the proposed nonlinear output, the limitation of input - output linearization, namely the singularity when $x_2$ is zero. Section seven is the well known linear state feedback, the advantage of this approach is the reduced number of parameters when compared with the neural network linearization scheme. Finally, conclusions are formulated in section eight. 

The overall analysis in this paper includes concepts such as, state variables, Taylor linearization, operating points, transfer function, natural frequency, input - output linearization, approximate linearization with neural network, approximate linearization with linear state feedback. It is important to mention that simulations are used to illustrate the main mathematical results so the reader can visualize different circuit dynamics when parameters or inputs are changed. 

\section{Nonlinear LC Circuit}
Consider the nonlinear circuit, Figure 1, with three elements in parallel, current source $u(t)$, capacitance $C_0$, and Josephson junction $L(x_1)$ \cite{devoret, wendin}. 

\begin{figure}[ht]
  \centering
  \includegraphics[scale=1.0]{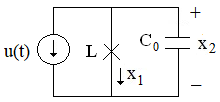}
  \caption{Nonlinear circuit with Josephson junction, $x_1$ is the inductance current and $x_2$ is the capacitance voltage.}
\end{figure}

The nonlinear inductance $L(x_1)$ follows the equation,
\begin{align}
L(x_1(t))= \frac{L_0}{ \sqrt {1 - (\frac{x_1(t)}{I_0})^2}}, \hspace{3 mm} L_0 = \frac{\kappa}{I_0}.
\end{align}

Where $I_0$ is the Josephson junction critical current and $\kappa$ is a constant, in the simulations below $\kappa = 1$.

From the circuit,
\begin{align}
L(x_1) \dot {x}_1 = x_2,\\
C_0 \dot {x}_2 = -(u + x_1).
\end{align}

The proposed output $y$ is a nonlinear function of the state variables $(x_1, x_2)$, for example,
\begin{align}
y = \frac{1}{2}L(x_1)x_1^2 + \frac{1}{2}C_0 x_2^2,
\end{align}

the dot in the dynamic equations means time derivative.

\section{Taylor linearization}
A linear version of the nonlinear LC circuit can be formulated for an operating point by using Taylor linearization \cite{close}. Writing the circuit equations again,
\begin{align}
\dot {x}_1 &= f_1(x_1, x_2, u) = \frac{1}{L(x_1)} x_2,\\
\dot {x}_2 &= f_2(x_1, x_2, u) = -\frac{1}{C_0}x_1 - \frac{1}{C_0}u,\\
y &= h(x_1, x_2, u) = \frac{1}{2}L(x_1)x_1^2 + \frac{1}{2}C_0 x_2^2.
\end{align}

The input $u(t)$ is periodic with frequency $\omega$ plus a constant term,
\begin{align}
u(t) &= a_0 + a_1 \sin{\omega t}\\
\bar{u} &= a_0.
\end{align}

Finding the equilibrium point $(\bar{x}_1, \bar{x}_2)$,
\begin{align}
\frac{1}{L(\bar{x}_1)} \bar{x}_2 = 0,\\
-\frac{1}{C_0} \bar{x}_1 - \frac{1}{C_0}\bar{u} = 0. 
\end{align}

Solving,
\begin{align}
\bar{x}_1 = - \bar{u},\\
\bar{x}_2 = 0. 
\end{align}

The linearized system $(z_1, z_2)$ at the equilibrium point ${(\bar{x}_1, \bar{x}_2, \bar{u})}$ is calculated as follows,
\begin{align}
\begin{bmatrix}\dot{z}_1 \\
\dot{z}_2\\ 
\end{bmatrix} = 
\begin{bmatrix}\frac{\partial f_1}{\partial x_1} & \frac{\partial f_1}{\partial x_2}\\
\frac{\partial f_2}{\partial x_1} & \frac{\partial f_2}{\partial x_2} \\ 
\end{bmatrix}_{(\bar{x}_1, \bar{x}_2, \bar{u})}
\begin{bmatrix}z_1 \\
z_2\\ 
\end{bmatrix} + 
\begin{bmatrix}\frac{\partial f_1}{\partial u} \\
\frac{\partial f_2}{\partial u}\\ 
\end{bmatrix}_{(\bar{x}_1, \bar{x}_2, \bar{u})} u \\
y_0 = 
\begin{bmatrix}\frac{\partial h}{\partial x_1} & \frac{\partial h}{\partial x_2}
\end{bmatrix}_{(\bar{x}_1, \bar{x}_2, \bar{u})}
\begin{bmatrix}z_1 \\
z_2\\ \end{bmatrix} + \begin{bmatrix}\frac{\partial h}{\partial u} \end{bmatrix}_{(\bar{x}_1, \bar{x}_2, \bar{u})} u.
\end{align}

Finally, the linearized system at the equilibrium point  $(\bar{x}_1, \bar{x}_2, \bar{u})$,
\begin{align}
\begin{bmatrix}\dot{z}_1 \\
\dot{z}_2\\ 
\end{bmatrix} = 
\begin{bmatrix}0 & \frac{1}{L(\bar{x}_1)}\\
- \frac{1}{C_0} & 0 \\ 
\end{bmatrix}
\begin{bmatrix}z_1 \\
z_2\\ 
\end{bmatrix} + 
\begin{bmatrix}0 \\
- \frac{1}{C_0}\\ 
\end{bmatrix} u \\
y_0 = 
\begin{bmatrix} c_{11} & c_{12} \end{bmatrix}
\begin{bmatrix}z_1 \\
z_2\\ \end{bmatrix}.
\end{align}

Where the output parameters are,
\begin{align}
c_{11}= L(\bar{x}_1)\bar{x}_1\Big[\frac{1}{2I_0L_0^2}L^2(\bar{x}_1)\bar{x}_1^2 + 1\Big], \hspace{3 mm}c_{12} = 0,
\end{align}
and $y_0$ is the Taylor linearized output. 

It is also possible to define another output $y_l$ for the linearized system as a nonlinear function of the new state variables $(z_1, z_2)$, for example,
\begin{align}
y_l = \frac{1}{2}L(\bar{x}_1)z_1^2 + \frac{1}{2}C_0 z_2^2.
\end{align}

To summarize there are three outputs: $y(t)$ is the nonlinear output of the nonlinear circuit, $y_0(t)$ is the linear output for the Taylor linearized circuit, $y_l(t)$ is the nonlinear output for the Taylor linearized circuit.

To visualize the dynamics of both, nonlinear system and linearized system, simulations for two different operating points were carried out in octave \cite{octave} with sampling period T = 0.01 s and a total number of samples NS = 2000. 

The parameters for the first operating point are, $a_0 = 0.05$, $a_1 = 0$, $I_0 = 0.2 A$, $C_0 = 0.1 F$. Figure 2 shows the state variables $(x_1, x_2)$ for the nonlinear system versus the state variables $(z_1, z_2)$ for the Taylor linearized system, and Figure 3 presents the outputs $\{y(x_1, x_2), y_0(z_1, z_2), y_l(z_1, z_2)\}$. Notice that for this operating point $(0.05, 0)$ the Taylor approximation is close to the actual nonlinear system, the operating point $\bar{x}_1$ is smaller in magnitude compared with the critical current $I_0$ of the Josephson junction.

\begin{figure}[H]
  \centering
  \includegraphics[scale=0.8]{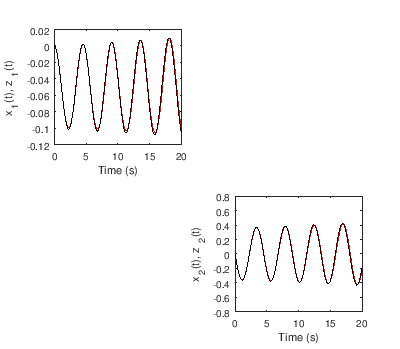}
  \caption{Nonlinear system (red) versus Taylor linearization, $\bar{u} = 0.05$.}
\end{figure}

\begin{figure}[H]
  \centering
  \includegraphics[scale=0.7]{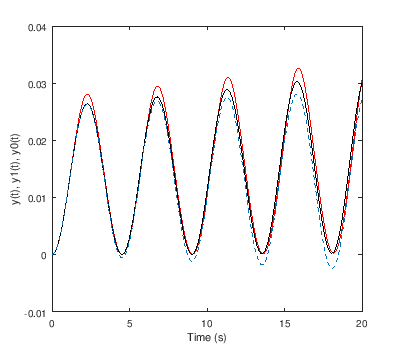}
  \caption{Nonlinear output $y$ (red), Taylor linearization output $y_0$ (dashed), nonlinear output from linearized circuit $y_l$ (black), $\bar{u} = 0.05.$}
\end{figure}

The parameters for the second operating point are, $a_0 = 0.1$, $a_1 = 0$, $I_0 = 0.2 A$, $C_0 = 0.1 F$. Figure 4 shows the state variables $(x_1, x_2)$ for the nonlinear system versus the state variables $(z_1, z_2)$ for the linearized system, and Figure 5 presents the outputs $(y(x_1, x_2), y_0(z_1, z_2), y_l(z_1, z_2))$, for this operating point $(0.1, 0)$ the approximation error increases, $\bar{x}_1$ is closer to $I_0$.

Notice, Figures 2 and 4, the reduction in frequency when $\bar{x}_1$ increases its magnitude from 0.05 to 0.1.

\begin{figure}[H]
  \centering
  \includegraphics[scale=0.8]{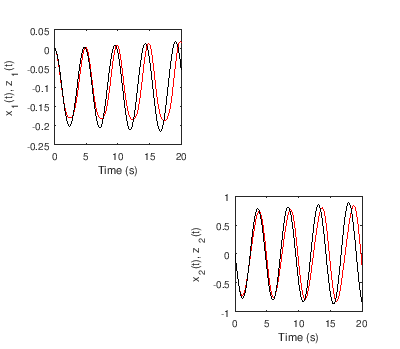}
  \caption{Nonlinear system (red) versus Taylor linearization, $\bar{u} = 0.1$.}
\end{figure}

\begin{figure}[H]
  \centering
  \includegraphics[scale=0.7]{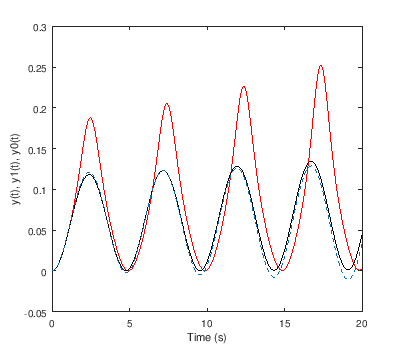}
  \caption{Nonlinear output $y$ (red), Taylor linearization output $y_0$ (dashed), nonlinear output from linearized circuit $y_l$ (black), $\bar{u} = 0.1$.}
\end{figure}

\section{Transfer function}
Calculating the transfer function for the Taylor linearized system,
\begin{align}
T(s) = \frac{k_0}{s^2 + \frac{1}{L_0 C_0}\{1 - (\frac{\bar{x}_1}{I_0})^2\}^\frac{1}{2}}, \hspace{3 mm} k_0 =\frac{c_{11}}{C_0^2},  
\end{align}
with s the Laplace operator. Solving for the natural frequency,
\begin{align}
\omega_0 = \frac{1}{\sqrt{L_0 C_0}}\{1 - (\frac{\bar{x}_1}{I_0})^2\}^\frac{1}{4},
\end{align}
notice that $\omega_0$ is a function of the operating point $\bar{x}_1$. Figure 6 illustrates the change in the natural frequency as function of the operating point $\bar{x}_1$, increasing the magnitude of $\bar{x}_1$ decreases $\omega_0$.

\begin{figure}[H]
  \centering
  \includegraphics[scale=0.7]{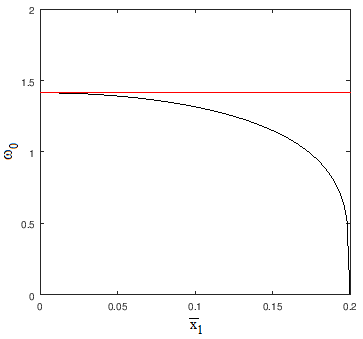}
  \caption{Natural frequency $\omega_0$ of the linearized circuit as function of the operating point $\bar{x}_1$.}
\end{figure}

\section{Exact feedback linearization}
For the original nonlinear circuit it is possible to obtain a linear dynamics between a new input $v(t)$ and the nonlinear output $y(t)$ by calculating a linearizing feedback $u(t) = f(x_1, x_2, v)$. This procedure is an exact linearization also know as input - output linearization, the linearizing input u(t) cancels the nonlinear terms exactly and produces a linear differential equation between y(t) and v(t) \cite{isidori}.

The output from the nonlinear circuit,
\begin{align}
y = h(x_1, x_2) = \frac{1}{2}L(x_1)x_1^2 + \frac{1}{2}C_0 x_2^2,
\end{align}

taking the time derivative,
\begin{align}
\dot{y} = \frac{\partial h(x_1, x_2)}{\partial x_1} \dot{x}_1 + \frac{\partial h(x_1, x_2)}{\partial x_2} \dot{x}_2,
\end{align}

where,
\begin{align}
\dot {x}_1 &= \frac{1}{L(x_1)} x_2,\\
\dot {x}_2 &= -\frac{1}{C_0}x_1 - \frac{1}{C_0}u,
\end{align}

calculating,
\begin{align}
\frac{\partial h(x_1, x_2)}{\partial x_1} &= x_1 L(x_1) \Big [1 + \frac{1}{2I_0L_0^2}x_1^2 L^2(x_1) \Big]\\ 
\frac{\partial h(x_1, x_2)}{\partial x_2} &= C_0 x_2.
\end{align}

Replacing all terms for $\dot{y}$ results,
\begin{align}
\dot{y} = \frac{1}{2I_0L_0^2} x_1^3 x_2 L^2(x_1) - x_2 u.
\end{align}

Then the input u,
\begin{align}
u = \frac{1}{x_2} \Big(\tau y - v + \frac{1}{2I_0L_0^2} x_1^3 x_2 L^2(x_1) \Big),\\
x_2 \neq 0,
\end{align}

cancels the nonlinear terms and produces a linear dynamics between the new input $v(t)$ and the output $y(t)$,
\begin{align}
\dot{y} + \tau y = v.
\end{align}

The linearization is exact but, in this case for the given output, there is a problem because the feedback u(t) works only when $x_2 \neq 0$.

\section{Neural network feedback linearization}
Figure 7 is a block diagram that illustrates an approach to approximate a linear dynamics between an external input v(t) and the nonlinear circuit output y(t) using machine learning to obtain the linearizing feedback u(t).  

The reference model, in this case $\tau = 1$, is a first order system,
\begin{align}
\dot{y} + \tau y = v.
\end{align}

The block NN known as neural network is a parametric model that can be trained to minimize a performance index $J$ by adjusting its parameters \cite{norgaard}. The neural network has three inputs $(x_1, x_2, v)$ and one output $u$, in other words, the network learns the map $u(t) = \hat{f}(x_1, x_2, v)$ to approximate the linear dynamics of the reference model by minimizing the index J.

For a discrete time system the performance index J can be,
\begin{align}
J = \sum_{k = 0}^{NS} (y_d(k) - y(k))^2
\end{align}
where NS is the number of samples and k the simulation time. In the simulations below the sampling time is T = 0.01 s and NS = 1000.

\begin{figure}[H]
  \centering
  \includegraphics[scale=1.0]{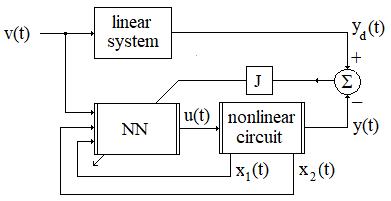}
  \caption{Feedback linearization with neural network, the network parameters are adjusted to minimize the performance index J}
\end{figure}

Figure 8 shows a neural network with architecture $NN(3, 8, 1)$, three inputs, N = 8 neurons in the hidden layer, and one output. 

The neural network equations are,
\begin{align}
u &= \sum_{i = 1}^{N} c_i h_i\\
u &= \sum_{i = 1}^{N} c_i \sigma(\omega_{i1}x_1 + \omega_{i2}x_2 + \omega_{i3}v)
\end{align}

The training algorithm \cite{bremermann} finds the parameters $(\omega_{ij}, c_i)$ by minimizing the performance index J for a given input v(t). 

\begin{figure}[H]
  \centering
  \includegraphics[scale=0.8]{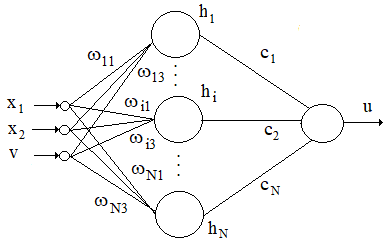}
  \caption{Multilayer neural network to learn the map $u = \hat{f}(x_1, x_2, v)$ by minimizing a performance index J, the number of neurons in the hidden layer is N.}
\end{figure}

Figure 9 shows a successful training example with final J = 0.0068, the nonlinear circuit output $y(t)$ follows the desired linear dynamics $y_d(t)$. Figure 10 presents the external input v(t) and the linearizing feedback $u(t) = \hat{f}(x_1, x_2, v)$, there are sudden changes in $u(t)$ at the discontinuity points of $v(t)$, the feedback is bounded $\vert u(t) \vert < I_0$. 

Table 1 is the final set of parameters $(\omega_{ij}, c_i)$ for the $NN(3,8,1)$ and performance index J = 0.0068.

\begin{figure}[H]
  \centering
  \includegraphics[scale=0.7]{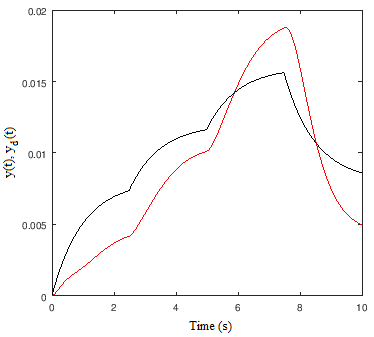}
  \caption{Feedback linearization with NN(3,8,1), nonlinear circuit output $y(t)$ (red) and desired output $y_d(t)$, final performance index J = 0.0068.}
\end{figure}

\begin{figure}[H]
  \centering
  \includegraphics[scale=0.8]{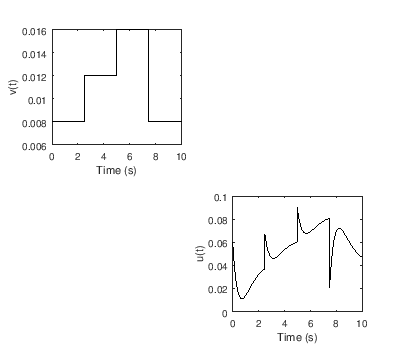}
  \caption{Feedback linearization with NN(3,8,1), reference input v(t) and linearizing input $u(t) = \hat{f}(x_1, x_2, v)$.}
\end{figure}

\begin{table}[H]
\caption{Neural network parameters (3,8,1): $\omega_{ij}$ input weights, $c_i$ output weights, $i = 1, 2, ..., N$.}
\centering
 \begin{tabular}{||c c c c||} 
 \hline
 $\omega_{i1}$ & $\omega_{i2}$ & $\omega_{i3}$ & $c_i$ \\ [0.5ex] 
 \hline\hline
-0.7958 & -1.4315 & 0.2044 & 1.2613 \\ 
 \hline
-0.8271 & -0.0502 &  -1.4862 &  -3.7185\\
 \hline   
1.7226 &  0.3998 &   2.7922 &   0.4928\\
 \hline
-0.0450 &  1.1977 &  0.5819 &  0.4234\\
 \hline
0.5180 &  -0.9837 &  -0.7239 &  -0.3991\\
 \hline
-1.0230 &  0.1191 &  -0.2298 &  1.2199\\
 \hline
0.8017 &  0.3706 &  1.0007 & -0.2855\\
 \hline
-1.2453 &  1.2427 &   0.6089 &  0.6291 \\ [1ex] 
 \hline
\end{tabular}
\end{table}

\section{Linear state feedback}
Figure 11 is a block diagram that shows how to approximate a linear dynamics between an external input v(t) and the nonlinear circuit output y(t) using a linear state feedback u(t). The main advantage here compared with the neural network approach is the reduction in both, the number of parameters and the time required to find the final values. However, it is clear that a linear state feedback is not capable of cancelling nonlinear terms.  

The linear state feedback equation is,
\begin{align}
u = k_1 x_1 + k_2 x_2 + k_3 v
\end{align}

The same training algorithm \cite{bremermann} finds the parameters $(k_1, k_2, k_3)$ by minimizing the performance index J for a given input v(t). 

\begin{figure}[H]
  \centering
  \includegraphics[scale=1.0]{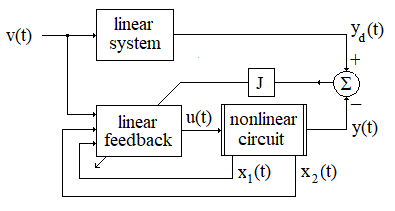}
  \caption{Linear state feedback, the parameters are adjusted to minimize J.}
\end{figure}

Figure 12 shows a successful training example with final performance index J = 0.00487, the nonlinear circuit output $y(t)$ follows the desired linear dynamics $y_d(t)$. Figure 13 presents the external input v(t) and the linear state feedback $u(t)$, the feedback is bounded $\vert u(t) \vert < I_0$. 

The final linear state feedback,
\begin{align}
u = -0.6176 \hspace{1 mm} x_1 + 0.0410 \hspace{1 mm} x_2 + 1.8195 \hspace{1 mm} v
\end{align}

\begin{figure}[H]
  \centering
  \includegraphics[scale=0.7]{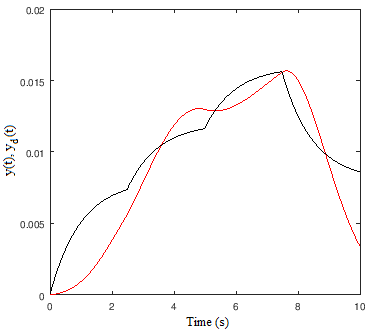}
  \caption{Linear state feedback, actual output (red) and desired output, final performance index J = 0.00487.}
\end{figure}

\begin{figure}[H]
  \centering
  \includegraphics[scale=0.8]{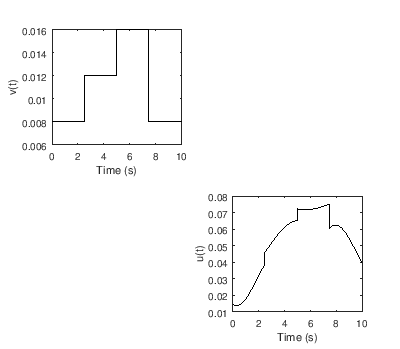}
  \caption{Linear state feedback, external input v and state feedback u.}
\end{figure}

\section{Conclusions}
Superconducting circuits with Josephson junctions are the building blocks of quantum bits or qubits. The main motivation for this paper was to study, from a classical point of view, a nonlinear LC circuit with a Josephson junction as the nonlinear inductance, the analysis integrates circuit and control theories.

Taylor linearization is a good approximation for the nonlinear circuit when the operating point is small in magnitude compared with the critical current of the junction, circuit outputs can be linear or nonlinear functions of the state variables.

Feedback input - output linearization is an alternative to Taylor linearization by cancelling exactly the nonlinear terms between a new input and the output, but as shown here sometimes there are limitations due to singularities in the control law. 

Machine learning, with neural networks, can be used as parametric nonlinear control law to approximate an exact input - output feedback linearization, this approach can overcome to some degree the problem of singularities but the number of parameters and training the network can be computationally intensive. Finally a linear feedback with trainable parameters can be used to approximate a linear dynamics for a known input sequence, the linear feedback has a reduced number of parameters when compared with the neural network but is not capable of cancelling nonlinear terms. In both learning settings, linear and nonlinear feedback, the control law is bounded $\vert u(t) \vert < I_0$ to avoid singularities, this limitation also impacts the approximation of the desired linear dynamics.

\section*{Acknowledgments}
Alberto Delgado thanks his employer the National University of Colombia for the support during the sabbatical year 2018. Also, the author thanks Prof. Pierre Rouchon and the organizers of the program \textit{Measurement and Control of Quantum Systems: Theory and Experiments} at the Institute Henri Poincare - Paris, for hosting an academic visit from 16 April - 13 July, 2018.

\end{document}